\documentclass[11pt, a4paper]{article}

\usepackage{jheppub}
\usepackage{amsmath, braket, orcidlink}

\newcommand{\EE}{\mathrm{EE}}

\preprint{OU-HET 1323}

\title{
Entanglement entropy minimization and global symmetry violation in scatterings
}

\author{
Shinya Kanemura\,\orcidlink{0000-0001-6819-4057}\,$^{a}$ and 
Masanori Tanaka\,\orcidlink{0000-0002-1303-7043}\,$^{b}$
}

\emailAdd{kanemu@het.phys.sci.osaka-u.ac.jp}
\affiliation{$^{a}$Department of Physics, The University of Osaka, Toyonaka, Osaka 560-0043, Japan}

\emailAdd{tanaka@pku.edu.cn}
\affiliation{$^{b}$Center for High Energy Physics, Peking University, Beijing 100871, China}

\abstract{The concept of quantum information has provided new ways to investigate the theoretical structure of particle interactions. In particular, it has been shown that extrema of quantum entanglement in scattering and decay processes can be related to symmetries and characteristic properties of particles. We focus on a more specific question: whether the minimization of entanglement entropy generated in scattering processes is systematically related to the suppression of interactions that violate global symmetries or selection rules. We conjecture that the minimization of the entanglement entropy can select the symmetry-preserving point when a new interaction opens a symmetry-violating scattering channel. We investigate the plausibility of this conjecture by considering interactions beyond the Standard Model (BSM) that violate global symmetries or the corresponding selection rules, including lepton number violation, baryon number violation, lepton flavor violation, and flavor-changing neutral currents. We show that, when symmetry-violating interactions open final-state sectors that are orthogonal to the symmetry-preserving sector, the symmetry-preserving point becomes a local minimum of the entanglement entropy. Our results support the possibility that the minimization of entanglement provides a common information-theoretic principle underlying the suppression of BSM phenomena relevant to symmetry-violating interactions.}

\begin{document}
\maketitle

\section{Introduction}

Despite the remarkable success of the Standard Model (SM), the origin of the strong suppression of symmetry-violating interactions remains an open question.
A number of processes that violate approximate symmetries or selection rules, including lepton-number and flavor violations, have been tightly constrained by experiments.
Such suppressions are usually attributed to a high scale of new physics beyond the SM (BSM), small couplings, or special structures in the interaction parameters.
It is therefore intriguing to ask whether there exists a more general principle that selects symmetry-preserving interactions.

Quantum information provides a novel perspective on this question.
Scattering processes naturally generate quantum correlations among the kinematic and internal degrees of freedom of particles~\cite{Lello:2013bva,Seki:2014cgq,Peschanski:2016hgk}.
Recent studies have shown that extrema of quantum-information quantities, such as entanglement entropy (EE), entanglement power, and quantum magic, can be associated with characteristic structures of particle interactions.
In particular, entanglement suppression has been related to emergent or enhanced symmetries in various physical systems~\cite{Beane:2018oxh,Low:2021ufv,Beane:2021zvo,Bai:2022hfv,Liu:2022grf,Liu:2023bnr,Ghosh:2023rpj,Carena:2023vjc,Kowalska:2024kbs,Hu:2024hex,McGinnis:2025brt,Busoni:2025dns,Hu:2025lua,McGinnis:2025xgt,McGinnis:2025iab,Low:2026evp,Low:2026kxb,Yamagishi:2026ihf}.
Related connections between entanglement minimization and flavor or CP structures have also been explored~\cite{Quinta:2022sgq,Thaler:2024anb,Li:2026udy}.
Conversely, entanglement maximization has been associated with gauge structures and symmetry-enhanced points~\cite{Cervera-Lierta:2017tdt,Nunez:2025xds,Miller:2023ujx,Carena:2025wyh,Liu:2025pny,Liu:2025iwh,Li:2026kha,Cao:2026aye}.
These observations suggest that extrema of quantum-information quantities may encode characteristic properties of fundamental interactions, although the physical origin of such extremization remains an open question.

Motivated by these developments, we investigate the conjecture that the entanglement minimization can provide a general criterion for suppressing BSM phenomena relevant to symmetry-violating interactions.
We consider the SM as a reference theory in which a certain symmetry or selection rule is exactly or approximately respected, and introduce an additional interaction that violates this structure.
When the new interaction opens an additional final-state sector that is distinguishable from the symmetry-preserving sector in the reduced Hilbert space, the interference contribution between the two sectors vanishes in the reduced density matrix.
In this situation, the symmetry-preserving point becomes a local minimum of the EE.
Therefore, the minimum-entanglement condition favors the suppression of the symmetry-violating amplitude of scattering processes.
The orthogonality condition provides a sufficient criterion for the above argument, while more general situations depend on the overlap between the reduced density matrices of the two sectors.

We examine this conjecture in several representative BSM phenomena; i.e., lepton-number violation, baryon-number violation, lepton-flavor violation, and flavor-changing neutral currents. 
Although the microscopic origins of these phenomena are different, they share a common feature: the symmetry-violating interaction opens a final-state sector with a different quantum-number structure from the symmetry-preserving one.
This common Hilbert-space structure leads to a similar form of the EE despite the different microscopic origins of these BSM phenomena.
We demonstrate that the minimum EE condition selects the symmetry-preserving limit when the corresponding final-state sectors satisfy the required orthogonality condition.
Our results suggest that the suppression of global symmetry-violating phenomena can be understood from a unified viewpoint based on quantum information theory.

This paper is organized as follows.
In Section~\ref{sec:EE_scattering}, we introduce the formalism for defining EE in scattering processes.
In Section~\ref{sec:general_minEE}, we derive the general relation between EE minimization and symmetry-preserving limits.
Sections~\ref{sec:LNVs}--\ref{sec:FCNC} apply this framework to representative examples of symmetry-violating interactions.
Finally, we discuss the implications and limitations of our results in Section~\ref{sec:discussion} and summarize our conclusions in Section~\ref{sec:conclusion}.

\section{Entanglement entropy in scattering processes} 
\label{sec:EE_scattering}

In this section, we define the EE in two-body scattering processes.
In general, the total Hilbert space for two-body scatterings between particles $a$ and $b$ can be given by~\cite{Fan:2017hcd, Low:2024hvn}
\begin{align}
\mathcal{H}_{\rm tot} 
= \bigotimes_{i = a,b} \left( \mathcal{H}_{\rm kin}^{i} \otimes \mathcal{H}_{\lambda}^{i}  \right) \,,
\end{align}
where $\mathcal{H}_{\rm kin}^{i}$ denotes the Hilbert space for momentum states and $\mathcal{H}_{\lambda}^{i}$ represents other quantum states labeled by $\lambda_{i}$ such as spin, particle type, and other inner degrees of freedom (i.e. color states). 

The initial state is taken as a product state $\ket{\rm in} = \ket{p_a,p_b} \ket{\lambda_a,\lambda_b}$ which does not contain entanglement between the quantum degrees of freedom considered here.
Considering the final state after a scattering process from the initial state, the outgoing state can be expressed as~\cite{Low:2024hvn, Liu:2025pny}
\begin{align}
\label{eq:final_state}
\ket{\rm fin}
=
\sum_{\lambda_c,\lambda_d}
\int d\Pi_{c} d\Pi_{d}\,
\mathcal{M}_{\lambda_a \lambda_b \lambda_c \lambda_d}
(p_a,p_b,p_c,p_d)
\ket{p_c,p_d}
\ket{\lambda_c,\lambda_d} \,.
\end{align}
The Lorentz-invariant phase-space measure $d\Pi_{i}$ is defined by
\begin{align}
d\Pi_i \equiv \frac{d^3 p_i}{(2\pi)^3 2E_i} \,,
\end{align}
where $E_{i}$ denotes the energy of the particle $i$. 
The states $\ket{p_i}$ and $\ket{\lambda_i}$ denote the kinetic and internal quantum states defined in $\mathcal{H}_{\rm kin}^{i}$ and $\mathcal{H}_{\lambda}^{i}$, respectively. 
The scattering amplitude is defined through the $T$ matrix as
\begin{align}
\braket{p_c,p_d;\lambda_c,\lambda_d|T|p_a,p_b;\lambda_a,\lambda_b}
=
i(2\pi)^4
\mathcal{M}_{\lambda_a\lambda_b\lambda_c\lambda_d}
(p_a,p_b,p_c,p_d)
\delta^{4}(p_a+p_b-p_c-p_d),
\end{align}
where $\delta^{4}(\cdots)$ represents the delta function for energy-momentum conservation. 
In the following analysis, we focus on the entanglement generated by the scattering process and therefore omit the non-interacting contribution $\mathbf{1}$ in the $S$ matrix $S=\mathbf{1}+iT $.

For the kinetic states, we assume that they are experimentally measured by an appropriate way to avoid quantum entanglement from the momentum states~\cite{Seki:2014cgq,Faleiro:2016lsf,Grignani:2016igg,Aguilar-Saavedra:2025njw, Fabbrichesi:2026jvm}. 
This means that we use the post-measurement state for kinetic states. 
The projection operator characterizing the wave function collapse due to measurements can be defined by~\cite{Low:2024hvn, Thaler:2024anb}
\begin{align}
\Pi_{p_{3}p_{4}} 
 =\ket{p_{3}, p_{4}} \bra{p_{3}, p_{4}} \otimes \left[ 
\sum_{\lambda_{3}, \lambda_{4} } \ket{\lambda_{3}, \lambda_{4}} \bra{\lambda_{3}, \lambda_{4}}
\right] \,, 
\end{align}
where the first part plays a role to fix the kinetic states. 
Using this projection operator, the post-measurement out state with fixed kinetic states can be given by 
\begin{align}
\label{eq:out_state}
\ket{\rm out} = \frac{ \Pi_{p_{3} p_{4}} \ket{\rm fin} }{ \sqrt{ \braket{ {\rm fin}|\Pi_{p_{3} p_{4}} | {\rm fin} }} } \,. 
\end{align}
We then define the density matrix after scatterings by 
\begin{align}
\rho_{f} = \ket{\rm out} \bra{\rm out}  \,.
\end{align}

Considering quantum correlations, we should specify which two subsystems we focus on.
In the following, we decompose the total Hilbert space into the bipartite system $\mathcal{H}_{A}$ and $\mathcal{H}_{B}$ as $\mathcal{H}_{\rm tot} = \mathcal{H}_{A} \otimes \mathcal{H}_{B}$. 
We then define the reduced density matrix for the subsystem $A$ as 
\begin{align}
\rho_{A} = {\rm tr}_{\mathcal{H}_{B}} \left[ \rho_{f} \right] \,,
\end{align}
where ${\rm tr}_{\mathcal{H}}[\cdots]$ means tracing out all quantum states in a Hilbert space $\mathcal{H}$. 
In our analysis, we use the linear entropy as a measure of the quantum entanglement. 
The linear entropy is defined by~\cite{Zanardi:2000zz}
\begin{align}
\label{eq:LinearEE}
\EE = 1 - {\rm tr}_{\mathcal{H}_{A}} \left[ (\rho_{A})^2 \right] \,. 
\end{align}
In the subsequent sections, we use this entropy in discussing the information-theoretic implications of phenomena beyond the SM. 
For the definition of the bipartite system, we should choose the most appropriate one for each specific problem. 
Therefore, we define them in each section later.

\section{Global symmetry violation and orthogonal final-state sectors}
\label{sec:general_minEE}

Before addressing specific BSM phenomena, we first discuss the general argument used in the subsequent sections.
In the SM, several accidental global symmetries exist, such as lepton-number conservation.
If new physics effects violate such symmetries, they can open scattering channels that are forbidden by the corresponding conservation laws or selection rules.

For instance, if lepton number is violated, one may consider the scattering processes
$\ket{W^{+}W^{+}} \to \ket{W^{+}W^{+}}~\text{or}~\ket{\ell^{+}\ell^{+}}$,
where $W$ and $\ell$ denote the $SU(2)_{L}$ gauge boson and charged lepton, respectively.
Although the second channel is absent in the SM, it can be induced by lepton-number-violating interactions in BSM models.
Thus, the symmetry-violating interaction connects final-state sectors that are separated in the SM.

To obtain a general argument, we consider an out state $\ket{\Psi}$ after scattering processes including symmetry-violating effects.
The state can be decomposed into the SM contribution $\ket{\psi_{\rm SM}}$ and the BSM contribution $\ket{\psi_{\rm BSM}}$ as
\begin{align}
\ket{\Psi} = \sqrt{1 - P} \ket{\psi_{\rm SM}} + e^{i \phi} \sqrt{P} \ket{\psi_{\rm BSM}} \,,
\end{align}
where $P$ and $\phi$ represent the relative weight of the two scattering sectors and their relative phase, respectively.

For this state, the reduced density matrix for the subsystem $A$ is given by
\begin{align}
\label{eq:rhoA_general}
\rho_{A} 
& = {\rm tr}_{\mathcal{H}_{B}} \left( \ket{\Psi} \bra{\Psi} \right) \nonumber \\
& = (1 - P) \rho_{A}^{\rm SM} + P \rho_{A}^{\rm BSM} 
+ \sqrt{P(1-P)}
\left[
e^{-i \phi} \rho_{A}^{\rm mix}
+ e^{i \phi} \left( \rho_{A}^{\rm mix} \right)^{\dagger}
\right] \,,
\end{align}
with
\begin{align}
\rho_{A}^{\rm SM/BSM}
=
{\rm tr}_{\mathcal{H}_{B}}
(\ket{\psi_{\rm SM/BSM}}\bra{\psi_{\rm SM/BSM}}),
\quad
\rho_{A}^{\rm mix}
=
{\rm tr}_{\mathcal{H}_{B}}
(\ket{\psi_{\rm SM}}\bra{\psi_{\rm BSM}}).
\end{align}

In the following discussion, we focus on cases where the reduced interference term between the SM and BSM sectors after tracing out states in the subsystem $B$ vanishes for an appropriate choice of bipartition $\rho_{A}^{\rm mix}=0$.
This condition is realized when the two sectors occupy different quantum-number sectors in the reduced Hilbert space.
Using Eq.~\eqref{eq:rhoA_general}, the linear entropy can then be expressed as
\begin{align}
\label{eq:EE_general}
\EE(\rho_{A}) = 1  - \left( 1 - P \right)^2 p_{\rm SM} - P^2 p_{\rm BSM} - 2 P (1-P) p_{\rm mix} \,,
\end{align}
where
\begin{align}
p_{\rm SM/BSM}
\equiv
\operatorname{tr}_{\mathcal{H}_{A}}
\left[
\left( \rho_{A}^{\rm SM/BSM} \right)^2
\right],
\quad
p_{\rm mix}
\equiv
\operatorname{tr}_{\mathcal{H}_{A}}
\left(
\rho_{A}^{\rm SM}\rho_{A}^{\rm BSM}
\right).
\end{align}
Considering the second derivative with respect to $P$ for the linear entropy in Eq.~\eqref{eq:EE_general}, we obtain
\begin{align}
\frac{\partial^2 \EE}{\partial P^2}
=
-2(p_{\rm SM}+p_{\rm BSM}-2p_{\rm mix})
=
-2\operatorname{tr}_{\mathcal{H}_{A}}
\left[
(\rho_{A}^{\rm SM}-\rho_{A}^{\rm BSM})^2
\right]
\leq 0 .
\end{align}
Therefore, the entropy minimum can only appear at the boundary points $P=0$ or $P=1$.
On the other hand, we obtain
\begin{align}
\label{eq:dEEdP}
\frac{\partial \EE}{\partial P}
=
2(p_{\rm SM}-p_{\rm mix})
-
2P\operatorname{tr}_{\mathcal{H}_{A}}
\left[
(\rho_{A}^{\rm SM}-\rho_{A}^{\rm BSM})^2
\right].
\end{align}
Thus, the point $P=0$ becomes a local minimum as long as $p_{\rm SM}>p_{\rm mix}$ is satisfied.

In addition, if $P=0$, we obtain $\EE(\rho_{A})=1-p_{\rm SM}$, whereas $\EE(\rho_{A})=1-p_{\rm BSM}$ is obtained for $P=1$.
Therefore, if $p_{\rm SM}>p_{\rm BSM}$ and $p_{\rm SM}>p_{\rm mix}$ are satisfied, the point $P=0$ gives the global minimum of the entropy.

We emphasize that the above argument relies on the requirement $\rho_{A}^{\rm mix}=0$.
This condition is naturally realized in many symmetry-violating processes when the new interaction connects final states belonging to different quantum-number sectors.
However, symmetry violation itself is not sufficient to guarantee this requirement; the relation depends on the Hilbert-space structure of the scattering process and the choice of bipartition.
Therefore, the minimum-entanglement criterion applies to a class of symmetry-violating phenomena in which the corresponding final-state sectors satisfy this orthogonality condition.

This finding does not contradict the relation between maximal entanglement and global symmetry found in Ref.~\cite{Carena:2025wyh}.
In their analysis, maximal entanglement selects special relations among nonzero Higgs self-couplings and thereby identifies a symmetry-enhanced point in the parameter space.
Our analysis instead considers a different situation in which a reference theory, which corresponds to the SM in our scenario, respects a global selection rule at tree level while additional interactions act as symmetry-violating sources.
When such interactions open new final-state sectors for which the interference term vanishes in the reduced density matrix, the entropy minimization favors the symmetry-preserving limit.
Thus, the results in Ref.~\cite{Carena:2025wyh} and our results correspond to different mechanisms through which symmetry-related structures can emerge from extrema of quantum-information quantities.

Our analysis is also related to the discussions in Refs.~\cite{Cervera-Lierta:2017tdt,Nunez:2025xds,Yamagishi:2026ihf}.
Their works connect gauge symmetries with maximal entanglement and minimal quantum magic in conventional Yang-Mills theories.
In contrast, our analysis focuses on scattering processes with new interactions that violate global symmetries that connect otherwise separated final-state sectors.
The difference between gauge and global symmetries may therefore provide an interesting direction for understanding why different symmetry structures are associated with different entanglement extrema.

\section{Lepton number violation} \label{sec:LNVs}

We first apply the discussion in Section~\ref{sec:general_minEE} to the LNV processes.
We consider the following two scattering channels:
\begin{align}
\label{eq:LNVprocess}
W^{-}(p_{1}, \lambda_{1}) +  W^{-}(p_{2}, \lambda_{2}) \to W^{-}(p_{3}, h_{3}) + W^{-}(p_{4}, h_{4}) ~~ \text{or} ~~ \ell_{i}^{-}(p_{3}, h_{3}) + \ell_{j}^{-} (p_{4}, h_{4}) \,.
\end{align}
The second channel changes the total lepton number by two units $\Delta L = 2$ and constitutes the electroweak subprocess underlying neutrinoless double-beta decay~\cite{deGouvea:2013zba}.
Therefore, we expect that the EE defined in these scattering channels is significant in discussing the relation between the quantum entanglement and LNV signals. 

The KamLAND-Zen Collaboration reports the lower limit $T_{1/2}^{0\nu}>3.8\times10^{26}\,{\rm yr}$ at the $90\%$ confidence level for the neutrinoless double-beta decay of ${}^{136}{\rm Xe}$ \cite{KamLAND-Zen:2024eml}.
The absence of an observed signal thus motivates us to investigate whether the minimum-entanglement criterion favors suppression of the corresponding LNV amplitude.

We note that the superposition of the two channels in Eq.~\eqref{eq:LNVprocess} does not conflict with the exact superselection rule of the lepton number. 
Both final states carry the same total electric charge, while lepton number is not an exact conserved charge in the presence of the $\Delta L=2$ interaction under consideration.
Consequently, the $WW$ and $\ell\ell$ components are not separated by an exact lepton-number superselection rule.

Because identical particles can appear in the final state, the subsystems $A$ and $B$ should not be interpreted as intrinsic particle labels. 
Instead, we define them in terms of the two asymptotic outgoing momentum modes as 
\begin{align}
A\equiv p_{3} \,, \quad B \equiv p_{4} \,. 
\end{align}
The required Bose or Fermi exchange symmetry is understood to be included in the corresponding scattering amplitudes. 
This momentum-mode definition provides an unambiguous bipartition even when the two outgoing particles belong to the same species.

For fixed initial helicities $(\lambda_{1},\lambda_{2})$, the normalized
out-state is written as
\begin{align}
\begin{aligned}
\ket{{\rm out}}_{\lambda_{1}\lambda_{2}}
=
\frac{1}{\sqrt{\mathcal{N}_{\lambda_{1}\lambda_{2}}}}
\Bigg[
&
\sum_{\lambda_{3},\lambda_{4}} \mathcal{M}^{WW}_{\lambda_{1}\lambda_{2};\lambda_{3}\lambda_{4}} \ket{W_{\lambda_{3}}^{-}}_{3} \ket{W_{\lambda_{4}}^{-}}_{4} \\ 
&+ \sum_{i,j} \sum_{h_{3},h_{4}} \mathcal{M}^{\ell\ell}_{\lambda_{1}\lambda_{2};ij,h_{3}h_{4}} 
\ket{\ell_{i,h_{3}}^{-}}_{3} \ket{\ell_{j,h_{4}}^{-}}_{4}
\bigg] \,,
\end{aligned}
\label{eq:LNVoutstate}
\end{align}
where $\mathcal{M}^{WW}_{\lambda_{1}\lambda_{2};\lambda_{3}\lambda_{4}}$ and $\mathcal{M}^{\ell\ell}_{\lambda_{1}\lambda_{2};ij,h_{3}h_{4}}$ denote the amplitudes for the lepton-number-conserving (LNC) and LNV channels, respectively.
The subscripts $3$ and $4$ label the outgoing momentum modes.

The normalization factor $\mathcal{N}_{\lambda_{1}\lambda_{2}}$ is given by
\begin{align}
\mathcal{N}_{\lambda_{1}\lambda_{2}} = \mathcal{N}_{WW}^{\lambda_{1}\lambda_{2}} + \mathcal{N}_{\ell \ell}^{\lambda_{1}\lambda_{2}} \,, 
\end{align}
with 
\begin{align}
\mathcal{N}_{WW}^{\lambda_{1}\lambda_{2}} = \sum_{\lambda_{3}, \lambda_{4}} |\mathcal{M}_{\lambda_{1}\lambda_{2}; \lambda_{3} \lambda_{4}}^{WW}|^2 \,, \quad 
\mathcal{N}_{\ell \ell} = \sum_{i,j} \sum_{h_{3}, h_{4}} |\mathcal{M}_{\lambda_{1},\lambda_{2};h_{3}h_{4}}^{\ell_{i} \ell_{j}}|^2 \,.
\end{align}

As a bipartite system to estimate the quantum entanglement, we focus on the one-particle Hilbert spaces associated with the two momentum modes
\begin{align}
\mathcal{H}_{\rm tot} = \mathcal{H}_{A} \otimes \mathcal{H}_{B} \,,  \quad \mathcal{H}_{A} = \mathcal{H}_{A}^{W} \oplus \mathcal{H}_{A}^{\ell} \,, \quad
\mathcal{H}_{B} = \mathcal{H}_{B}^{W} \oplus \mathcal{H}_{B}^{\ell} \,,
\end{align}
with 
\begin{align}
\mathcal{H}_{A/B}^{W} = {\rm span} \left\{ \ket{W_{\lambda}^{-}}_{A/B} \right\}_{\lambda = 0, \pm} \,, \quad  \mathcal{H}_{A/B}^{\ell} = {\rm span} \left\{ \ket{\ell_{i, \lambda}^{-}}_{A/B} \right\}_{i= e, \mu, \tau \, ;\, \lambda = \pm} \,,
\end{align}
where the label $\lambda$ denotes the helicity eigenvalue for each particle. 

In general, we can select different definitions of the bipartite system in calculating the EE as discussed in Ref.~\cite{Low:2024hvn}. 
We note that our results can be changed if we choose a different bipartite system as we mentioned before. 
However, the choice of the bipartite system that divides the total Hilbert space into the individual one-particle sectors seems natural in two-body scattering processes.

Tracing out the quantum state for the particle $B$, the reduced density matrix for the particle $A$ is given by
\begin{align}
\label{eq:rhoA_LNV}
\rho_{A}^{\lambda_{1}\lambda_{2}} 
= {\rm tr}_{\mathcal{H}_{B}} \left[ \ket{\rm out}_{\lambda_{1} \lambda_{2}} \bra{\rm out}_{\lambda_{1} \lambda_{2}} \right] 
= P_{WW} \rho_{A}^{WW} \oplus P_{\ell \ell} \rho_{A}^{\ell \ell} \,,
\end{align}
with
\begin{align}
& P_{WW} = \frac{\mathcal{N}_{WW}}{ \mathcal{N}_{WW} + \mathcal{N}_{\ell \ell}} \,,  \quad 
P_{\ell \ell} = \frac{\mathcal{N}_{\ell \ell}}{ \mathcal{N}_{WW} + \mathcal{N}_{\ell \ell}} \,, \\ 
& \left( \rho_{A}^{WW}\right)_{\lambda_{1} \lambda_{2}} = \frac{1}{\mathcal{N}_{WW}} \sum_{\lambda_{3}} \mathcal{M}_{\lambda_{1}\lambda_{3}}^{WW} \left( \mathcal{M}^{WW}_{\lambda_{2} \lambda_{3}} \right)^{*} \,, \\ 
& \left( \rho_{A}^{\ell \ell}\right)_{(i_{1},\lambda_{1} )(i_{2}, \lambda_{2})} = \frac{1}{\mathcal{N}_{\ell \ell}} \sum_{i_{3},\lambda_{3}} \mathcal{M}_{i_{1} i_{3}; \lambda_{1}\lambda_{3}}^{\ell \ell} \left( \mathcal{M}^{\ell \ell}_{i_{2} i_{3}; \lambda_{2} \lambda_{3}} \right)^{*} \,.
\end{align}

We note that the off-diagonal terms between the $WW$ and $\ell\ell$ sectors vanish after tracing out the quantum state for the particle $B$ because the one-particle states of a $W$ boson and a charged lepton are orthogonal.
This implies that $\rho_{A}^{\rm mix} = 0$ and $p_{\rm mix} = 0$ are valid in this case. 
As a result, the reduced density matrix has a block-diagonal structure and the corresponding linear entropy can be given by
\begin{align}
\EE = 1 - P_{WW}^2 {\rm tr}\left[ (\rho_{A}^{WW})^2 \right] - P_{\ell \ell}^2 {\rm tr}\left[ (\rho_{A}^{\ell \ell})^2 \right] \,.
\end{align}

For an illustrative analytic limit, we additionally focus on the longitudinal final-state channel $\ket{W_{L}^{-}}_{3}\ket{W_{L}^{-}}_{4}$.
This situation is related to the result with the high-energy limit~\cite{Lee:1977eg}. 
Under this assumption, we obtain
\begin{align}
\EE = 1 - (1 - P_{\ell \ell})^2 - P_{\ell \ell}^2 {\rm tr} \left[ (\rho_{A}^{\ell \ell})^2 \right] \,.
\end{align}

We then apply this result to a specific scenario to clarify its implications.
We here consider a case where the LNV process is induced via the Weinberg operator~\cite{Weinberg:1979sa}. 
This operator gives the vertex for the Nambu-Goldstone boson $w^{\pm}$ and the lepton as~\cite{Weinberg:1979sa, Fuks:2020zbm}
\begin{align}
\mathcal{L}_{\rm eff} =  \frac{C_{ij}}{\Lambda} (H \cdot \overline{L_{i}^{c}}) (L_{j} \cdot H) + {\rm h.c.} ~\ni~ \frac{C_{ij}}{\Lambda} w^{-} w^{-} \overline{\ell_{i}^{c}} \,\ell_{j} \,.
\end{align}
At energies much larger than the EW scale, the corresponding amplitude satisfies the following relation because of the equivalence theorem~\cite{Lee:1977eg}
\begin{align}
\mathcal{M} \left( W_{L}^{-}W_{L}^{-}\to\ell_{i}^{-}\ell_{j}^{-} \right)
=
\mathcal{M}
\left( w^{-}w^{-}\to\ell_{i}^{-}\ell_{j}^{-} \right)
+ \mathcal{O} \, \left(\frac{m_{W}}{\sqrt{s}}\right) \,. 
\label{eq:equivalenceLNV}
\end{align}
Therefore, the amplitude for the process $W_{L}W_{L} \to \ell \ell$ in the high-energy limit can be given by 
\begin{align}
\mathcal{M}_{ij}^{\ell \ell} = \frac{C_{ij}}{\Lambda} K(s, \theta) \,,
\end{align}
where $K(s, \theta)$ characterizes the kinematic dependence from the spinor algebra. 

On the other hand, the Weinberg operator gives the Majorana neutrino mass matrix after the EW symmetry breaking as~\cite{Weinberg:1979sa}
\begin{align}
\label{eq:mnu_Cij}
(m_{\nu})_{ij} = C_{ij} \frac{v^2}{2\Lambda} \,.
\end{align}
Moreover, since we take the high-energy limit, the produced charged leptons via the Weinberg operator are left-handed. 
Therefore, the amplitude for the process $W_{L}W_{L} \to \ell \ell$ can be expressed in terms of the neutrino mass matrix as 
\begin{align}
\label{eq:M_ll_mnu}
\mathcal{M}_{ij,--}^{\ell \ell} = \mathcal{A}(s, \theta) \left( {m_{\nu}} \right)_{ij} \,, \quad 
\mathcal{M}_{ij,++}^{\ell \ell} = \mathcal{M}_{ij,-+}^{\ell \ell} = \mathcal{M}_{ij,+-}^{\ell \ell} = 0 \,,
\end{align}
where $A(s, \theta)$ represents the kinematic factor, which does not affect the final expression of the EE.
The label $\pm$ denotes the helicity states for charged leptons in the out state. 
Combining Eq.~\eqref{eq:M_ll_mnu} with Eq.~\eqref{eq:rhoA_LNV}, the linear entropy can take a simple form
\begin{align}
\label{eq:EE_LNV}
\EE = 2 P_{\ell \ell} \left( 1- P_{\ell \ell} \right) + P_{\ell \ell}^2 \left[ 1- \frac{\sum_{i} m_{\nu_i}^4}{(\sum_{\nu_i}m_{\nu_i}^2)^2} \right] \,,
\end{align}
where $m_{\nu_{i}}$ denotes the mass eigenvalue of neutrinos obtained by diagonalizing Eq.~\eqref{eq:mnu_Cij}. 

Requiring the minimization of this entropy, we find two solutions:
\begin{align}
P_{\ell \ell} = 0 \quad \text{or} \quad P_{\ell \ell} = 1 ~~ \text{with} ~~ {\rm rank}(m_{\nu}) = 1 \,.
\end{align}
The second scenario indicates that only one neutrino has a nonzero mass. 
Taking into account neutrino oscillation results~\cite{Esteban:2024eli,Capozzi:2025bcr}, this solution is not plausible. 
Therefore, we can conclude that the first solution, which indicates the absence of the LNV processes, is a plausible prediction from the minimization of EE. 
This is entirely consistent with the conclusion obtained in Section~\ref{sec:general_minEE}.

Moreover, for the Weinberg operator, we obtain
\begin{align}
P_{\ell \ell} \propto \mathcal{N}_{\ell\ell}
=
\left|\mathcal{A}(s,\theta)\right|^{2} {\rm tr} \, \left(m_{\nu}m_{\nu}^{\dagger}\right) \,. 
\end{align}
If the kinematic factor $\mathcal{A}$ takes a nonzero value, the condition $P_{\ell\ell}=0$ therefore requires
\begin{align}
m_{\nu}=0
\quad \text{or} \quad
\frac{C}{\Lambda}=0 \,.
\end{align}
Thus, exact minimization does not predict small but nonzero Majorana neutrino masses. 
Instead, it selects the lepton-number-conserving limit in which the Weinberg operator and the corresponding $\Delta L=2$ amplitude vanish.

We note that our conclusion is not changed even if we do not take the high-energy limit. 
In the high-energy limit, we obtain Eq.~\eqref{eq:EE_LNV} which corresponds to a case $p_{\rm SM} = 1$ and $p_{\rm mix} = 0$ in Eq.~\eqref{eq:EE_general}. 
If the high-energy limit is not taken, in general we obtain $p_{\rm SM} \neq 1$ but $p_{\rm mix} = 0$ is still valid. 
Even if we consider such a case, the entropy still has a local minimum at the point $P_{\ell\ell}=0$ because $p_{\rm SM}>0$ always holds.

We emphasize that the EE considered in this work is defined for fixed kinematics. 
Therefore, the minimum entropy condition should not necessarily be interpreted as a requirement at all energy scales. 
Rather, it may select specific relations among amplitudes at characteristic kinematic points.
As we briefly discuss in subsection~\ref{sec:BSM_kinetic}, this energy scale dependence may explain why small BSM effects can appear at low-energy scales at which realistic experiments are performed.

\section{Baryon number violation} 
\label{sec:BNVs}

Proton decay is one of the characteristic signals of grand unified theories (GUTs)~\cite{Pati:1973rp}. 
A representative decay mode is $p \to \pi^{0} e^{+}$, which can be induced by baryon-number-violating interactions involving quarks and leptons~\cite{Langacker:1980js,Takhistov:2026aln}. 
The Super-Kamiokande Collaboration gives the lower bound $\tau(p \to \pi^{0} e^{+})>2.4\times10^{34}\,\text{years}$ at the $90\%$ confidence level~\cite{Super-Kamiokande:2020wjk}. 
In the EFT approach, this corresponds to that the GUT scale is significantly high. 
We here discuss the interpretation of the suppression of proton decays focusing on the minimum-entanglement criterion. 

We focus on the following partonic scattering processes related to baryon-number violation:
\begin{align}
\label{eq:GUT_process}
u(p_{1},h_{A})+u(p_{2},h_{B})
\to
u(p_{3},h_{C})+u(p_{4},h_{D})
~~\text{or}~~
\bar d(p_{3}/p_{4},h_{C})+e^{+}(p_{4}/p_{3},h_{D}) \,.
\end{align}
The first channel conserves baryon number and is allowed in the SM, while the second channel violates baryon and lepton numbers and can be induced by the same type of interactions that lead to proton decay. 
Since identical particles appear in the first final state, we distinguish the two subsystems by their outgoing momentum states. 
We identify subsystem $A$ with $\ket{p_{3}}$ and subsystem $B$ with $\ket{p_{4}}$.

The out state for the scattering processes in Eq.~\eqref{eq:GUT_process} can be written as
\begin{align}
\begin{aligned}
\ket{\rm out}_{ij}
=
\frac{1}{\sqrt{\mathcal N_{ij}}}
& \left[
\sum_{h_A,h_B}
\mathcal M^{uu}_{ij,h_Ah_B}
\ket{u_{i,h_A}}_{3}
\ket{u_{j,h_B}}_{4} \right. \\
& \left. +
\sum_{h_C,h_D}
\mathcal M^{\bar d e}_{ij,h_Ch_D}
\ket{\bar d_{h_C}}_{3}
\ket{e^{+}_{h_D}}_{4}
+
\sum_{h_C,h_D}
\mathcal M^{e\bar d}_{ij,h_Ch_D}
\ket{e^{+}_{h_D}}_{3}
\ket{\bar d_{h_C}}_{4}
\right] \,.
\end{aligned}
\end{align}
Here, $\ket{\cdots}_{3}$ and $\ket{\cdots}_{4}$ denote the states associated with the two outgoing momentum modes. 
For simplicity, color indices are not shown explicitly. 
They can be included in the amplitudes and density matrices without changing the following argument.
We define the normalization factors as
\begin{align}
\begin{aligned}
&\mathcal N_{ij}=N_{\rm SM}+N_{\rm BNV} \,, \\
&N_{\rm SM}
=
\sum_{h_A,h_B}
\left|
\mathcal M^{uu}_{ij,h_Ah_B}
\right|^{2} \,,
\quad
N_{\rm BNV}
=
\sum_{h_C,h_D}
\left(
\left|
\mathcal M^{\bar d e}_{ij,h_Ch_D}
\right|^{2}
+
\left|
\mathcal M^{e\bar d}_{ij,h_Ch_D}
\right|^{2}
\right) \,.
\end{aligned}
\end{align}

For the $uu$ final state, the notation $\ket{u}_{3}\ket{u}_{4}$ should be understood as the corresponding two-fermion Fock state. 
The Fermi statistics and the exchange contributions for the identical quarks are included in the usual definition of the scattering amplitude.

The reduced density matrix in this scenario also takes the block-diagonal form
\begin{align}
\label{eq:rhoA_GUT}
\rho_A
=
(1-P_{\rm BNV})\rho_A^{\rm SM}
\oplus
P_{\rm BNV}\rho_A^{\rm BNV} \,.
\end{align}
Since the SM and BNV final states contain different particle species in both subsystems, the Hilbert spaces for each one-particle state are orthogonal. 
Therefore, the interference term $\rho_A^{\rm mix}$ vanishes after tracing out quantum states in the subsystem $B$. 
In addition, the reduced density matrices for the subsystem $A$ have a block-diagonal structure because the particle species in the SM and BNV branches are different and hence $p_{\rm mix}=0$. 

The normalized density matrices for the two branches are given by
\begin{align}
\rho_A^{\rm SM}
&=
\frac{1}{N_{\rm SM}}
M_uM_u^{\dagger} \,,
\quad
\rho_A^{\rm BNV}
=
\frac{1}{N_{\rm BNV}}
\left(
M_{\bar d e}M_{\bar d e}^{\dagger}
\oplus
M_{e\bar d}M_{e\bar d}^{\dagger}
\right) \,,
\end{align}
where $\mathcal M^{X}_{ij,h_Ah_B}=(M_X)_{h_Ah_B}$. 
The weights of the two branches are defined by
\begin{align}
P_{\rm SM}
=
\frac{N_{\rm SM}}{\mathcal N_{ij}} \,,
\quad
P_{\rm BNV}
=
\frac{N_{\rm BNV}}{\mathcal N_{ij}} \,,
\quad
P_{\rm SM}+P_{\rm BNV}=1 \,.
\end{align}

Substituting Eq.~\eqref{eq:rhoA_GUT} into the definition of the linear entropy in Eq.~\eqref{eq:LinearEE}, we obtain
\begin{align}
\label{eq:EE_GUT}
\EE
=
1
-
(1-P_{\rm BNV})^{2}
{\rm tr}
\left[
(\rho_A^{\rm SM})^{2}
\right]
-
P_{\rm BNV}^{2}
{\rm tr}
\left[
(\rho_A^{\rm BNV})^{2}
\right] \,.
\end{align}
This expression has the same form as the result obtained for the LNV processes. 
In particular, in the case with $P_{\rm BNV} \ll 1$, the first derivative of the entropy at $P_{\rm BNV}=0$ is positive as confirmed in Eq.~\eqref{eq:dEEdP}. 
Therefore, the baryon-number-conserving point is a local minimum of the EE, and the minimum-entanglement criterion favors
\begin{align}
P_{\rm BNV}=0 \,.
\end{align}
Thus, the minimum-entanglement criterion favors the suppression of the baryon- and lepton-number-violating amplitudes related to proton decay. 

A simple realization is provided by leptoquarks that admit both quark--lepton and diquark couplings. 
When we use the notation in Ref.~\cite{Dekens:2018bci}, integrating out a heavy leptoquark scalar $S_{1}$ gives baryon-number-violating amplitudes proportional to $(z_{RR})_{11}(y_{RR})_{11}/M_{S_1}^{2}$, where $(y_{RR})_{11}$ and $(z_{RR})_{11}$ denote its quark-lepton and diquark couplings for the first generation fermion, respectively.
Our result indicates that the minimal entanglement conjecture favors $(z_{RR})_{11}(y_{RR})_{11}=0$ or the decoupling limit $M_{S_{1}} \to \infty$. 
Therefore, the absence of proton decays can be interpreted as the consequence of the minimal EE requirement.

\section{Lepton flavor violation} \label{sec:LFVs}

We next consider LFV processes.
Charged-lepton flavor violation (cLFV) provides an important probe of BSM, as such processes are highly suppressed in the SM and can be strongly enhanced by new interactions~\cite{Petcov:1976ff}.
From the viewpoint of the present work, LFV processes also provide a useful test of the general minimum-entanglement argument. 
As we show below, flavor-exchange processes realize the locally orthogonal Hilbert-space structure discussed in Section~\ref{sec:general_minEE}.
On the other hand, in several LFV channels, the minimization of the EE cannot predict their suppression when the bipartite system we used. 
The limitation of our analysis is discussed in Section~\ref{sec:discussion} in detail.

LFV decay processes such as $\mu \to e \gamma$, $\mu \to \bar ee e$, $\tau \to \mu \gamma$, $\tau \to \bar \mu \mu \mu$ are strongly constrained by various experiments. 
The strongest current limits on branching ratios for each process at the $90\,\%$ confidence level are given by ${\rm BR}(\mu \to e \gamma) < 1.5 \times 10^{-13}$~\cite{MEGII:2025gzr}, ${\rm BR}(\mu \to \bar e e e) < 1.0 \times 10^{-12}$~\cite{SINDRUM:1987nra}, ${\rm BR}(\tau \to \mu \gamma) < 9.5 \times 10^{-8}$~\cite{Belle-II:2026usi}, and ${\rm BR}(\tau \to \bar \mu \mu \mu) < 1.9 \times 10^{-8}$~\cite{Belle-II:2024sce}, respectively. 
Other cLFV decay modes have also been experimentally explored (i.e., see Refs.~\cite{Heeck:2016xwg,COMET:2025sdw,ParticleDataGroup:2026mpi}).

We here focus on the following charged lepton scattering process instead of LFV decays
\begin{align}
\label{eq:lilj_LFV}
\ell_{i}^{-}(h_{A}, p_{1}) + \ell_{j}^{+} (h_{B}, p_{2}) \to \ell_{i}^{-}(h_{C}, p_{3}) + \ell_{j}^{+} (h_{D}, p_{4}) ~~ \text{or} ~~ \ell_{j}^{-}(h_{C}, p_{3}) +  \ell_{i}^{+}(h_{D}, p_{4}) \,, 
\end{align}
where the labels $i$ and $j$ characterize the lepton flavor ($i\,,j = e \,, \mu \,, \tau$ with $i \neq j$). 
The first process denotes the lepton flavor-conserving (LFC) processes that occur within the SM. 
The second channel corresponds to the LFV process induced by new physics effects~\cite{Calibbi:2017uvl}. 

The second process in Eq.~\eqref{eq:lilj_LFV} can be predicted in new physics models such as extensions of the SM containing doubly-charged scalar bosons~\cite{Halprin:1982wm}, Majorana neutrinos~\cite{Halprin:1982wm}, or neutral pseudoscalar bosons~\cite{Hou:1995dg}. 
For instance, the LFV scattering process $e^{-} \mu^{+} \to e^{+} \mu^{-}$ is constrained by the measurement of muonium-antimuonium conversion probability at the Paul Scherrer Institute~\cite{Willmann:1998gd}. 
The expected future sensitivities of searches for muonium-antimuonium conversion probability measurements are discussed in Ref.~\cite{Fukuyama:2021iyw}.

Using Eq.~\eqref{eq:out_state}, the final state after the scatterings in Eq.~\eqref{eq:lilj_LFV} with the initial flavor state $\ket{i, j}$ can be expressed by 
\begin{align}
\label{eq:out_state_LFV}
\ket{\rm out}_{ij} = \frac{1}{\sqrt{\mathcal{N}_{ij}}} \left[ \sum_{h_{A},h_{B}} \mathcal{M}_{ij,h_{A}h_{B}}^{\rm LFC} \ket{\ell_{i,h_{A}}^{-}} \ket{\ell_{j,h_{B}}^{+}} + \sum_{h_{C}, h_{D}}\mathcal{M}_{ij,h_{C}h_{D}}^{\rm LFV} \ket{\ell_{j,h_{C}}^{-}} \ket{\ell_{i,h_{D}}^{+}} \right] \,,
\end{align}
where $\mathcal{M}_{ij,h_{A}h_{B}}^{\rm LFC}$ and $\mathcal{M}_{ij,h_{A}h_{B}}^{\rm LFV}$ represent the amplitudes for the LFC and LFV processes, respectively. 
The indices $(i,j)$ and $h_{X}\,(X = A, B, C, D)$ identify the flavor and helicity states, respectively. 
The prefactor $\mathcal{N}_{ij}$ is the normalization condition given by 
\begin{align}
\mathcal{N}_{ij} = \mathcal{N}^{\rm LFC}_{ij} + \mathcal{N}^{\rm LFV}_{ij} \,, 
\end{align}
with 
\begin{align}
\mathcal{N}^{\rm LFC}_{ij} = \sum_{h_{A}, h_{B}} \left| \mathcal{M}_{ij;h_{A}h_{B}}^{\rm LFC} \right|^2 \,, \quad 
\mathcal{N}^{\rm LFV}_{ij} = \sum_{h_{C}, h_{D}} \left| \mathcal{M}_{ij;h_{C}h_{D}}^{\rm LFV} \right|^2 \,.
\end{align}

Employing Eq.~\eqref{eq:out_state_LFV}, we can define the density matrix for the out state as 
\begin{align}
\rho_{f}^{ij} = \ket{\rm out}_{ij} \bra{\rm out}_{ij} \,.
\end{align}

We identify particles with negative and positive charges as the particle $A$ and $B$, respectively. 
Then, we take the following bipartite system 
\begin{align}
\mathcal{H}_{\rm tot} = \mathcal{H}_{A} \otimes \mathcal{H}_{B} ~~ \text{with} ~~
\mathcal{H}_{A} = \mathcal{H}_{A}^{\ell_{i}} \oplus \mathcal{H}^{\ell_{j}}_{A} \,, ~~
\mathcal{H}_{B} = \mathcal{H}_{B}^{\ell_{i}} \oplus \mathcal{H}_{B}^{\ell_{j}}  ~~ (i \neq j)\,,
\end{align}
where
\begin{align}
\mathcal{H}_{A/B}^{\ell_{i}} = {\rm span} \left\{ \ket{\ell_{i, h}^{\pm}} \right\}_{i= e , \mu , \tau ;\, h = \pm} \,. 
\end{align}
Since $\braket{i|j} = 0 \, (i \neq j)$, the reduced density matrix for the subsystem $A$ takes a block-diagonal form as follows
\begin{align}
\rho_{A} = \left( 1 - P_{ij}^{\rm LFV} \right) \rho_{A}^{\rm LFC} + P_{ij}^{\rm LFV} \rho_{A}^{\rm LFV} \,, 
\end{align}
with 
\begin{align}
&P_{ij}^{\rm LFV}
=
\frac{\mathcal{N}_{ij}^{\rm LFV}}{\mathcal{N}_{ij}^{\rm LFC} + \mathcal{N}_{ij}^{\rm LFV}} \,, \quad 
P_{ij}^{\rm LFC}
=
1-P_{ij}^{\rm LFV} \,, \\
& \left( \rho_{A}^{\rm LFC} \right)_{hh'} = \frac{1}{\mathcal{N}_{ij}^{\rm LFC}} \sum_{k = \pm} \mathcal{M}_{ij, hk}^{\rm LFC} \left( \mathcal{M}_{ij, h'k}^{\rm LFC} \right)^{*} \,,  \\ 
& \left( \rho_{A}^{\rm LFV} \right)_{hh'} = \frac{1}{\mathcal{N}_{ij}^{\rm LFV}} \sum_{k = \pm} \mathcal{M}_{ij, hk}^{\rm LFV} \left( \mathcal{M}_{ij, h'k}^{\rm LFV} \right)^{*} \,.
\end{align}
As a result, the linear entropy defined in Eq.~\eqref{eq:LinearEE} is given by 
\begin{align}
\label{eq:EE_LFV}
\EE(\rho_{A}) = 1 - \left( 1 - P_{ij}^{\rm LFV} \right)^2 {\rm tr} \left[ (\rho_{A}^{\rm LFC})^2 \right] - \left(P_{ij}^{\rm LFV} \right)^2 {\rm tr} \left[ (\rho_{A}^{\rm LFV})^2 \right] \,. 
\end{align}

Assuming $P_{ij}^{\rm LFC} \gg P_{ij}^{\rm LFV}$ and expanding to linear order in $P_{ij}^{\rm LFV}$, we obtain 
\begin{align}
\EE (\rho_{A}) \simeq  \EE(\rho_{A}^{\rm LFC}) + 2 P_{ij}^{\rm LFV} {\rm tr} \left[ (\rho_{A}^{\rm LFC})^2 \right] \,. 
\end{align}
This expression is the same as Eq.~\eqref{eq:EE_general} with $p_{\rm mix} = 0$. 
Since the first term corresponds to the contribution from the SM sector to the EE, it can always appear regardless of the existence of LFV processes. 
Considering the first derivative with respect to $P_{ij}^{\rm LFV}$, we can show  
\begin{align}
\left. \frac{\partial \EE}{\partial P_{ij}^{\rm LFV}} \right|_{P_{ij}^{\rm LFV} = 0} = 2 {\rm tr} \left[ (\rho_{A}^{\rm LFC})^2 \right] \geq 0  \,. 
\end{align}
Therefore, we can conclude that the minimal entanglement entropy prefers
\begin{align}
\label{eq:minEE_flavor}
P_{ij}^{\rm LFV} = 0 \,. 
\end{align}
Since $P_{ij}^{\rm LFV} \propto \sum |\mathcal{M}_{ij}^{\rm LFV}|^2$, the minimal EE prefers the disappearance of the LFV processes. 
However, as we mentioned in Section~\ref{sec:LNVs}, the argument summarized in Eq.~\eqref{eq:minEE_flavor} should be understood as favoring the suppression of LFV amplitudes rather than requiring their exact absence.

When we obtain the result in Eq.~\eqref{eq:minEE_flavor}, we have focused only on the charged lepton scattering given in Eq.~\eqref{eq:lilj_LFV}. 
In addition to such processes, in general there are other scattering channels: 
\begin{itemize}
\item Processes in which only one lepton flavor changes (i.e., $\ket{e^{-} \mu^{+}} \to \ket{e^{-} \mu^{+}}$ or $\ket{\mu^{-} \mu^{+}}$)
\item Processes in which both lepton flavors change (i.e., $\ket{e^{-} \mu^{+}} \to \ket{e^{-} \mu^{+}}$ or $\ket{\tau^{-} \tau^{+}}$)
\end{itemize}
For the second category, we can obtain the same result by performing the similar calculation for the scattering process given in Eq.~\eqref{eq:lilj_LFV}. 
On the other hand, when only one of the two lepton flavors is changed, this local orthogonality is not generally satisfied, and the minimum-EE criterion does not in general lead to a model-independent suppression of the LFV amplitude.
We discuss this case in more detail in Section~\ref{sec:discussion}.

\section{Flavor-changing neutral currents} \label{sec:FCNC}

We next discuss the implications of the minimum-entanglement criterion for flavor-changing neutral-current (FCNC) processes.

In the SM, FCNC processes are absent at tree level and arise only through loop effects because of the Glashow-Iliopoulos-Maiani mechanism~\cite{Glashow:1970gm}. 
On the other hand, the FCNC processes can be induced by new physics effects at tree level. 
After the EW symmetry breaking, representative flavor-changing interactions with neutral bosons can be parametrized as
\begin{align}
\begin{aligned}
\mathcal{L}_{\rm FCNC}
={}&
-\frac{g}{2c_W} Z_\mu
\left[
\bar q_j\gamma^\mu
\left(
X_L^{ji}P_L+X_R^{ji}P_R
\right)q_i
+\text{h.c.}
\right]
\\
&- \sum_{\phi} \phi
\left[ 
\bar q_j \left( Y_{\phi L}^{ji}P_L+Y_{\phi R}^{ji}P_R \right) q_i + \text{h.c.}
\right]
+\cdots \,, 
\label{eq:FCNCLagrangian}
\end{aligned}
\end{align}
where $\phi$ denotes possible additional neutral scalar bosons.
The ellipsis represents other possible sources of FCNCs, such as effective four-quark interactions and loop-induced contributions.
In the following discussion, we do not specify the origin of the FCNC interaction and denote the corresponding flavor-changing amplitude by $\mathcal{M}_{\rm FCNC}$.

FCNC processes are also strongly constrained by several flavor experiments. 
For instance, in rare $K$ meson decays, the branching ratio ${\rm BR}(K^{+} \to \pi^{+} \bar \nu \nu) = 9.6^{+1.9}_{-1.8} \times 10^{-11}$ was measured~\cite{NA62:2026rwr}.
In rare $B$ meson decays, there are bounds ${\rm BR}(B^{0} \to \mu^{+} \mu^{-}) < 1.5 \times 10^{-10}$~\cite{CMS:2022mgd} and ${\rm BR}(B \to X_{s} \gamma) = (3.54 \pm 0.78 \pm 0.83) \times 10^{-4}$ with $E_{\gamma} > 1.8\,{\rm GeV}$, where $E_{\gamma}$ denotes the measured photon energy in the signal-$B$ meson rest frame~\cite{Belle-II:2022hys}. 
For other meson decays relevant to FCNCs, see Ref.~\cite{ParticleDataGroup:2026mpi}.

As a representative process, we consider
\begin{align}
q_i(p_{1}, h_{1}, a_1) + \bar q_j (p_{2}, h_{2}, a_2) \to 
\begin{cases}
q_i(p_{3}, h_{3}, a_3) + \bar q_j(p_{4}, h_{4}, a_4) \,, \\ 
q_j(p_{3}, h_{3}, a_3) + \bar q_i(p_{4},h_{4}, a_4) \,,
\end{cases}
\label{eq:FCNCprocess}
\end{align}
where $i\neq j$, and $q_i$ and $q_j$ denote either up-type or down-type quarks.
The labels $h_k$ and $a_k$ denote the helicity and color states, respectively.
Since the two outgoing particles are a quark and an antiquark, we identify them as subsystems $A$ and $B$, respectively.
The following analysis is similar to that for the LFV processes, while quarks have additional color degrees of freedom.

For fixed initial quantum numbers, the outgoing state can be written as
\begin{align}
\begin{aligned}
\ket{\mathrm{out}}
=
\frac{1}{\sqrt{\mathcal N}}
\sum_{h_3,h_4} \sum_{a_3,a_4}
\left[
\mathcal M_{\rm FC}^{a_3a_4;h_3h_4}
\ket{q_i,a_3,h_3} \ket{\bar q_j,a_4,h_4} \right. \\ \left.
+
\mathcal M_{\rm FCNC}^{a_3a_4;h_3h_4}
\ket{q_j,a_3,h_3} \ket{\bar q_i,a_4,h_4}
\right],
\end{aligned}
\label{eq:FCNCout}
\end{align}
where $\mathcal{M}_{\rm FC}$ denotes the flavor-conserving (FC) amplitude.
The normalization factor is given by
\begin{align}
\mathcal N=N_{\rm FC}+N_{\rm FCNC} \,, \quad 
N_X=
\sum_{a_3,a_4}
\sum_{h_3,h_4}
\left|
\mathcal M_X^{a_3a_4;h_3h_4}
\right|^2
\qquad
(X=\mathrm{FC},\mathrm{FCNC}) \,.
\end{align}

We consider the following bipartite system including the flavor, helicity, and color degrees of freedom
\begin{align}
\mathcal H_A
&=
\left(
\mathcal H_{q_i}\oplus\mathcal H_{q_j}
\right)
\otimes
\mathcal H_{\rm hel}^{q}
\otimes
\mathcal H_{\rm color}^{q} \,,
\nonumber\\
\mathcal H_B
&=
\left(
\mathcal H_{\bar q_j}\oplus\mathcal H_{\bar q_i}
\right)
\otimes
\mathcal H_{\rm hel}^{\bar q}
\otimes
\mathcal H_{\rm color}^{\bar q} \,.
\end{align}
Here, $\mathcal H_{q_i,\bar q_i}$, $\mathcal H_{\rm hel}^{q,\bar q}$, and $\mathcal H_{\rm color}^{q,\bar q}$ denote the Hilbert spaces for flavor, helicity, and color, respectively.
We work at the perturbative parton level and include the color states as internal degrees of freedom.

The reduced density matrix thus takes the block-diagonal form
\begin{align}
\rho_A
=
(1-P_{\rm FCNC})\rho_A^{\rm FC}
\oplus
P_{\rm FCNC}\rho_A^{\rm FCNC} \,,
\quad
P_{\rm FCNC}
=
\frac{N_{\rm FCNC}}{N_{\rm FC}+N_{\rm FCNC}} \,,
\label{eq:FCNCrho}
\end{align}
where $\rho_A^{\rm FC}$ and $\rho_A^{\rm FCNC}$ are normalized reduced density matrices including the color and helicity information of the corresponding branches.
QCD interactions can generate nontrivial correlations among the color and helicity states.
However, such correlations do not change the orthogonality between the flavor-conserving and FCNC branches because $\braket{q_i,a,h|q_j,a',h'}=0$ for $i\neq j$ in the SM.
The same relation holds for the antiquark states.
Therefore, the interference term introduced in Section~\ref{sec:general_minEE} vanishes after tracing over subsystem $B$, namely $\rho_A^{\rm mix}=0$.
In addition, the reduced density matrices for the FC and FCNC branches have the orthogonal flavor sector for the subsystem $A$, and hence $p_{\rm mix}=0$.

The linear entropy is then given by
\begin{align}
\EE
=
1
-
(1-P_{\rm FCNC})^2
\mathrm{tr}
\left[
(\rho_A^{\rm FC})^2
\right]
-
P_{\rm FCNC}^2
\mathrm{tr}
\left[
(\rho_A^{\rm FCNC})^2
\right] \,.
\label{eq:FCNCEE}
\end{align}
Focusing on the case with a small FCNC contribution $P_{\rm FCNC}\ll1$, we obtain
\begin{align}
\EE
\simeq
\EE(\rho_A^{\rm FC})
+
2P_{\rm FCNC}
\mathrm{tr}
\left[
(\rho_A^{\rm FC})^2
\right]
+
\mathcal O(P_{\rm FCNC}^2) \,,
\end{align}
and hence
\begin{align}
\left.
\frac{\partial \EE}{\partial P_{\rm FCNC}}
\right|_{P_{\rm FCNC}=0}
=
2\,
\mathrm{tr}
\left[
(\rho_A^{\rm FC})^2
\right]
>0 \,.
\label{eq:FCNCminimum}
\end{align}
Therefore, the flavor-conserving point is a local minimum of EE.
This result does not depend on the detailed color or helicity structure, because it follows from the orthogonality of the flavor states.

Thus, the minimum-entanglement criterion locally favors the suppression of the flavor-changing amplitude,
\begin{align}
\mathcal M_{\rm FCNC} \to 0 \,.
\end{align}
This conclusion can be applied to different sources of FCNCs, including flavor-changing $Z$ interactions and neutral-scalar-mediated FCNCs in extended Higgs sectors described by Eq.~\eqref{eq:FCNCLagrangian}.

If the loop-induced FCNC effect within the SM is included, $\mathcal M_{\rm FCNC}$ should be understood as the total flavor-changing amplitude of the SM and BSM contributions: $\mathcal{M}_{\rm FCNC} = \mathcal{M}_{\rm FCNC}^{\rm SM} + \mathcal{M}_{\rm FCNC}^{\rm BSM}$.
In this case, the condition in Eq.~\eqref{eq:FCNCminimum} implies that the minimal EE requirement predicts $\mathcal{M}_{\rm FCNC}^{\rm SM}(Q^{*}) + \mathcal{M}_{\rm FCNC}^{\rm BSM}(Q^{*}) = 0$ at a certain center-of-mass energy $Q^{*}$. 
However, this cancellation can be broken at low-energy scales satisfying $Q < Q^{*}$. 
As a result, the BSM contribution may be suppressed because of this requirement at low-energy scales.

\section{Discussion}
\label{sec:discussion}

We discuss the limitations and possible implications of the minimum-entanglement conjecture. 

\subsection{Robustness under the choice of entanglement measure}

Throughout this work, we have used the linear entropy defined in Eq.~\eqref{eq:LinearEE} as a measure of quantum entanglement.
It is natural to ask whether our conclusions depend on this particular choice.

Whenever $\rho_{A}^{\rm mix}=0$, the reduced density matrix takes the block-diagonal form $\rho_{A}=(1-P)\rho_{A}^{\rm SM}\oplus P\rho_{A}^{\rm BSM}$ used throughout Sections~\ref{sec:LNVs}--\ref{sec:FCNC}.
For such a block-diagonal state, the von Neumann entropy $S(\rho_{A})=-{\rm tr}_{\mathcal{H}_{A}}[\rho_{A}\ln\rho_{A}]$ admits the exact decomposition
\begin{align}
S(\rho_{A}) = H(P) + (1-P)\,S(\rho_{A}^{\rm SM}) + P\,S(\rho_{A}^{\rm BSM}) \,,
\end{align}
where $H(P)=-(1-P)\ln(1-P)-P\ln P$ is the Shannon entropy of the classical distribution that records which branch has occurred.
Differentiating with respect to $P$,
\begin{align}
\frac{\partial S}{\partial P} = \ln\frac{1-P}{P} - S(\rho_{A}^{\rm SM}) + S(\rho_{A}^{\rm BSM}) \,,
\end{align}
which diverges to $+\infty$ as $P\to0^{+}$, irrespective of the values of $S(\rho_{A}^{\rm SM})$ and $S(\rho_{A}^{\rm BSM})$.
The symmetry-preserving point is therefore not merely a local minimum but an infinitely steep one under the von Neumann entropy.
Therefore, opening any infinitesimal symmetry-violating channel increases the EE.
This confirms that the local-minimality conclusion reached throughout this work is not an artifact of the specific choice of the linear entropy as the entanglement measure.

We note that the minimal entropy conjecture mainly originates from the classical Shannon entropy $H(P)$ rather than from the internal entanglement of either branch individually.
However, as confirmed in Higgs decay analyses~\cite{Liu:2025iwh}, the classical entropy may also be important in discussing the relation between the extrema of entropy and the properties of particles. 
Our results support the claim that classical entropy, as well as quantum entropy, is an important quantity.

\subsection{Limitations of the minimum-entanglement argument}

As emphasized in Section~\ref{sec:LFVs}, there is a limitation of our argument on the minimal entanglement conjecture in LFV processes. 
We explain the details of the problem below. 

The relation between global symmetry violation (selection rule violation) and entropy minimization discussed in this work depends on the Hilbert-space structure of the relevant scattering channels.
In particular, local orthogonality between the symmetry-preserving and symmetry-violating branches provides a sufficient condition for $p_{\rm mix}=0$, and the symmetry-preserving point is then a local minimum of the EE.
However, this condition is not always satisfied.
A simple example is an LFV process in which only one of the two lepton flavors is changed.

For such an LFV process, the final state can be written in a simplified description including only the flavor states as
\begin{align}
\ket{\rm out} 
= A\ket{\mu^-\mu^+} + B\ket{e^-\mu^+} 
= \left( A\ket{\mu^-} + B\ket{e^-} \right) \otimes \ket{\mu^+} \,.
\end{align}
In this simple case, no flavor entanglement is generated even for a nonzero LFV amplitude.
Once the helicity degrees of freedom are included, such a factorization does not necessarily hold.
Nevertheless, in this class of processes, the flavor-conserving and flavor-violating branches are not locally orthogonal in both subsystems, and therefore $p_{\rm mix}$ does not generally vanish.
According to Eq.~\eqref{eq:dEEdP}, the condition for the flavor-conserving point to be a local minimum is $p_{\rm LFC}>p_{\rm mix}$.
Unlike the locally orthogonal case, this condition depends on the detailed structure of the amplitudes.
Therefore, the minimum-EE criterion does not give a model-independent suppression of LFV when only one of the two lepton flavors is changed.

Such processes can also be induced by the same LFV interactions that contribute to LFV decays such as $\mu\to e\gamma$.
It is therefore significant to investigate whether the entanglement in LFV decay processes provides additional information. 
Since such discussions may require more sophisticated analyses such as the three-body decay entanglement~\cite{Sakurai:2023nsc}, we leave this possibility for future work.

\subsection{Possible applications to other symmetries}

The general idea considered in this work may also be relevant to other symmetry-breaking phenomena.
For example, Ref.~\cite{Li:2026udy} showed that the quantum magic in hadron scattering has a local minimum at the CP-conserving point where the QCD $\theta$ term vanishes.
Although quantum magic is different from the EE, this result provides another example in which an extremum of a quantum-information quantity is related to a symmetry-preserving point.
It would therefore be significant to investigate whether a similar relation can be found between CP violation and scattering entanglement.

Chiral symmetry provides another useful example to discuss the scope of our argument.
Explicit chiral symmetry breaking interactions, such as fermion mass terms, connect left- and right-handed sectors in Hilbert spaces that are decoupled in the chiral limit.
When the corresponding chirality-violating scattering channel is orthogonal to the symmetry-preserving channel in a given bipartite system, our general argument implies that the chiral-symmetric limit can be locally favored by entropy minimization.
However, this conclusion should be distinguished from spontaneous chiral symmetry breaking.
In QCD, a nonzero quark condensate can arise even in the massless limit, while the Lagrangian remains chirally symmetric.
Therefore, the relation between entanglement extremization and spontaneous chiral symmetry breaking may require a different analysis.

\subsection{Appearance of small violation from the minimum-entanglement condition} \label{sec:BSM_kinetic}

An important implication of our result concerns the smallness of BSM effects at low energies.
In the previous sections, the EE is defined after fixing the outgoing momentum states.
Therefore, the condition $\mathcal M_{\rm BSM}=0$ obtained from the minimum-entanglement argument should be understood as a condition imposed at a certain physical scattering energy and kinematic configuration.
We discuss the possibility that the minimum-EE condition is realized at a characteristic energy scale $Q_{*}$.

If only a single effective operator contributes to the symmetry-violating amplitude, the condition $\mathcal M_{\rm BSM}(Q_{*})=0$ generally requires its Wilson coefficient to vanish.
In this case, the corresponding BSM effect also disappears at lower energies.
The situation can be qualitatively different when operators of different mass dimensions contribute to the same physical amplitude.
As a simple example, let us consider dimension-six and dimension-eight contributions with the same on-shell angular and internal-state structure as
\begin{align}
\mathcal M_{\rm BSM}^{\alpha}(Q,\theta)
=
f^{\alpha}(\theta)
\left[
C_{6}\frac{Q^{2}}{\Lambda^{2}}
+ C_{8}\frac{Q^{4}}{\Lambda^{4}}
\right] \,, 
\label{eq:minEE_dim6dim8}
\end{align}
where the label $\alpha$ collectively denotes helicity, flavor, color, and other internal quantum numbers.
The factors $C_{6}$ and $C_{8}$ denote the Wilson coefficients for each term. 
The common function $f^{\alpha}(\theta)$ in terms of the scattering angle $\theta$ is introduced only for simplicity and makes the cancellation independent of $\theta$ values and internal state.

Suppose that the minimum-EE condition is imposed at a certain center-of-mass energy $Q=Q_{*}$.
The vanishing of the symmetry-violating amplitude due to the minimal EE conjecture requires
\begin{align}
C_{6} = - C_{8}\frac{Q_{*}^{2}}{\Lambda^{2}} \,.
\label{eq:minEE_C6relation}
\end{align}
An important point is that this condition does not require either Wilson coefficient to vanish.
Instead, the minimum-entanglement condition fixes a relation between operators of different mass dimensions.
For $Q_{*}<\Lambda$ and $C_{8}=\mathcal O(1)$, Eq.~\eqref{eq:minEE_C6relation} implies
$|C_{6}|\sim Q_{*}^{2}/\Lambda^{2}\ll1$.
Thus, the coefficient of the leading dimension-six interaction is naturally suppressed by the minimum-EE condition.

The consequence becomes particularly transparent at energies below $Q_{*}$.
Using Eq.~\eqref{eq:minEE_C6relation}, the amplitude can be rewritten as
\begin{align}
\mathcal M_{\rm BSM}^{\alpha}(Q,\theta)
=
C_{8} f^{\alpha}(\theta)
\frac{Q^{2}}{\Lambda^{2}}
\left(
\frac{Q^{2}-Q_{*}^{2}}{\Lambda^{2}}
\right) .
\label{eq:minEE_lowenergy_amp}
\end{align}
Therefore, for $Q\ll Q_{*}$,
\begin{align}
\mathcal M_{\rm BSM}^{\alpha}(Q,\theta)
\simeq
-
C_{8} f^{\alpha}(\theta)
\frac{Q^{2}Q_{*}^{2}}{\Lambda^{4}} \,.
\label{eq:minEE_lowenergy_limit}
\end{align}
Compared with the naive dimension-six expectation
$\mathcal M_{6}\sim Q^{2}/\Lambda^{2}$ with an ${\cal O}(1)$ Wilson coefficient, the low-energy BSM amplitude contains an additional suppression factor $Q_{*}^{2}/\Lambda^{2}$.
Hence, although the BSM amplitude vanishes exactly at the minimum-entanglement scale $Q_{*}$, it can appear at lower energies with a suppressed magnitude.

This mechanism is not specific to dimension-six and dimension-eight operators.
More generally, a cancellation between operators of dimensions $d$ and $d+2$ at $Q_{*}$ can lead to the relation $C_{d}\sim -(Q_{*}^{2}/\Lambda^{2})C_{d+2}$, provided that the two operators contribute to the same on-shell amplitude structure.
The minimum-EE condition can therefore suppress the coefficient of the lower-dimensional operator, which would otherwise dominate the low-energy BSM signal.

This observation suggests a possible information-theoretic origin of the strong suppression of low-energy BSM phenomena.
Instead of requiring the leading BSM interaction to be absent, the minimum-entanglement condition at a characteristic scale may naturally select a small coefficient for the leading higher-dimensional operator through its relation to subleading operators.
A complete realization of this mechanism requires an understanding of what determines the characteristic scale $Q_{*}$ and whether the resulting relation among Wilson coefficients is stable in a UV-completed theory.
We leave these intriguing questions for future work.

\section{Conclusions} \label{sec:conclusion}

In this work, we have investigated the conjecture that the minimization of entanglement entropy (EE) in scattering processes can favor the suppression of interactions that violate global symmetries or selection rules.
We have shown that, when a symmetry-violating interaction opens a scattering channel that is orthogonal to the symmetry-preserving sector, the symmetry-preserving point becomes a local minimum of the EE.
This provides a simple general condition relating the minimization of entanglement to the suppression of symmetry-violating amplitudes.

We have examined the validity of this conjecture in several representative BSM phenomena such as lepton-number violation, lepton-flavor violation, flavor-changing neutral currents, and baryon-number violation. 
We have found that scattering processes relevant to these BSM phenomena have the required orthogonal Hilbert space structure.
In particular, for lepton-number violation induced by the Weinberg operator, the minimum-entanglement condition selects the lepton-number-conserving limit.
As a result, the suppression of scattering amplitudes relevant to these BSM phenomena is related to the minimization of EE. 
Although these BSM phenomena have different origins, the unified framework based on quantum information theory may provide a reason why these experimental signals do not appear at low-energy scales.

We have further discussed a possible implication for the smallness of BSM effects at low energies.
If the minimum-entanglement condition is realized at a characteristic energy scale, cancellations between operators with different mass dimensions can suppress the coefficient of the leading higher-dimensional interaction while allowing a small but nonzero BSM amplitude at lower energies.
This finding suggests a possible connection between the suppression of experimentally accessible BSM effects and the minimum-entanglement condition.

Our results support the possibility that the minimization of entanglement provides a common description of the suppression of interactions that violate global symmetries or selection rules.
Further studies are required to understand the physical origin of the characteristic scale at which the minimum-entanglement condition is realized and to determine how generally this idea can be extended to other symmetry structures.

\section*{Acknowledgements}

We are sincerely grateful to Qing-Hong Cao and the Center for High Energy Physics group at Peking University for their hospitality and support. 
This work was supported by JSPS under the Bilateral Program between Japan and China (JPJSBP120267416).

\bibliographystyle{JHEP}
\bibliography{reference} 


\end{document}